\documentstyle[aps,epsf,prl,floats]{revtex}

\def\sb#1{$_{#1}$}
\def\sp#1{$^{#1}$}

\begin{document}

\draft

\wideabs{

\title{Structural response to local charge order in underdoped but
superconducting La$\bf _{2-x}$(Sr,Ba)$\bf _x$CuO$\bf _4$ }

\author{M. Gutmann, E. S. Bo\v zin, and S. J. L. Billinge }
\address{Department of Physics and Astronomy and Center for
Fundamental Materials Research,\\ Michigan State University, 
East Lansing, MI 48824-1116.}

\date{\today}

\maketitle

\begin{abstract}
We report an anomalous local structural response in the CuO\sb{2}
planes associated with the appearance of charge inhomogeneities at low
temperature in underdoped but superconducting
La$_{2-x}$(Sr,Ba)$_x$CuO$_4$.  We used pair distribution function
analysis of neutron powder diffraction data. The increase in the Cu-O
bond length distribution at low temperature has an onset temperature
which correlates with observations of charge and spin freezing seen by
other probes.

\end{abstract}
}

Two unusual phenomena are observed in the normal state in the
underdoped cuprates: a pseudo-gap~\cite{timus;rpp99} in the
electronic~\cite{loese;s96,homes;prl93,renne;prl98,batlo;pc94} and
magnetic~\cite{warre;prl89,dai;prl96} densities of states, and the
possibility that the charge density in the superconducting planes of
these materials is inhomogeneously distributed, possibly in a striped
morphology~\cite{tranq;n95,yamad;prb98,mook;n99,schul;prl89,emery;pnas99,caste;zpb97,bianc;ssc97,white;prl98,marti;cm0300,phill;pmb99,marki;jpcs97}.
It is important to establish the role that charge inhomogeneities have
in the high-T\sb{c} phenomenon itself.  Unlike the pseudogap phenomenon, their
universal observation among different high-T$_c$ systems has not been
established. The strongest evidence for them in the cuprates is the observation
of {\it long range ordered} static charge stripes in
La$_{2-x-y}$Nd$_y$Sr$_x$CuO$_4$ compounds~\cite{tranq;n95}. These have
been seen in both insulating and superconducting compounds but they
appear to compete with the superconductivity~\cite{ichik;cm99}.  On
the other hand {\it fluctuating short range ordered} charge
stripes may play an active role in the high-T$_c$
phenomenon~\cite{emery;pnas99,caste;zpb97,bianc;ssc97,white;prl98,marti;cm0300,phill;pmb99,marki;jpcs97}. They
also give a natural explanation for the observation of incommensurate
spin fluctuations which have been seen in
La$_{2-x}$Sr$_x$CuO$_4$~\cite{yamad;prb98} and
YBa$_2$Cu$_3$O$_{6+\delta}$~\cite{mook;n00} as well as being able to
explain various other experimental
observations~\cite{mook;n99,ichik;cm99,mook;n00,hunt;prl99,lanza;jpcm99,petro;cm0300}. It
is important to establish both the ubiquity of charge inhomogeneities
in underdoped cuprates and their relationship to
superconductivity. Here we present diffraction evidence that
establishes the presence of temperature dependent atomic scale
structural inhomogeneities at low temperature in underdoped but
superconducting La$_{2-x}$(Sr,Ba)$_x$CuO$_4$ samples.  This
observation is naturally explained by the appearance of charge
inhomogeneities at low temperature in these samples. The
inhomogeneities appear at a temperature which correlates with spin and
charge freezing inferred from transport~\cite{ichik;cm99},
NQR~\cite{hunt;prl99}, and XANES~\cite{lanza;jpcm99} measurements.

We used the atomic pair distribution function (PDF)
analysis~\cite{egami;b;lsfd98} of neutron powder diffraction data to
study the local structure of La$_{2-x}$(Sr,Ba)$_x$CuO$_4$.  Structural
distortions coming from charge inhomogeneities appear in the PDF as an
anomalous broadening of the nearest neighbor in-plane Cu-O bond length
distribution~\cite{bozin;prl00}. The average in-plane Cu-O bond length
shortens on hole doping.  This is observed
experimentally~\cite{radae;prb94i} and is expected on the grounds that
holes are being doped into a $\sigma^*$ antibonding
band~\cite{goode;f92} thus stabilizing the bond.  Charge
inhomogeneities imply a coexistence of heavily and lightly doped
regions of the CuO\sb{2} plane.  The lattice will respond if the charge
inhomogeneities are fluctuating on phonon time-scales or slower.
This will result in a {\it distribution} of lengths for the in-plane
Cu-O bond and correspondingly to a {\it broadening} of the atomic pair
distribution. This can be measured directly using the PDF analysis of
neutron powder diffraction data. The PDF technique, which is common in
the study of glasses~\cite{egami;b;lsfd98}, is equally well applied to
crystalline systems where it reveals precise information about the
local atomic structure going beyond the approximation of
crystallinity~\cite{petko;prl99}.

\begin{figure}[tb]
\begin{center}$\,$
\epsfxsize=2.8in
\epsfbox{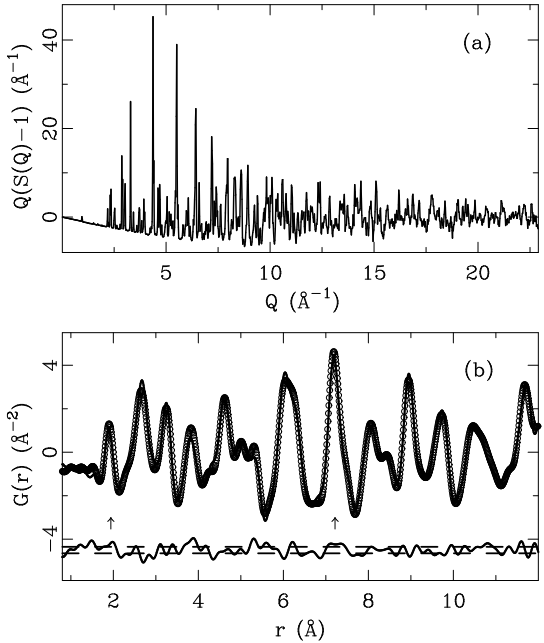}
\end{center}
\caption{(a) Reduced total scattering structure function, $Q[S(Q)-1]$, from 
La$_{1.875}$Sr$_{0.125}$CuO$_4$ at 300~K.  (b) The
resulting PDF, $G(r)$ (open circles). The solid line is a fit to the
data of the crystallographic model with the difference curve
below.  Arrows indicate the PDF peaks at $r = 1.9$~\AA\ and $r =
7.2$~\AA\ whose widths are plotted in Fig.~\ref{fig;u2vt}.  }
\protect\label{fig;data}
\end{figure}

Powdered samples of La$_{2-x}$(Sr,Ba)$_x$CuO$_4$ ($x=0.125,0.15$) of
$\sim 10$~g were synthesized using standard solid state
techniques~\cite{billi;prl93,breue;pc93}.  The samples were
characterized using x-ray diffraction and susceptibility measurements.
The oxygen content was verified by measuring the $c$-axis parameter
that was found to fall on the expected curve for stoichiometric
samples~\cite{radae;prb94i}.  Neutron powder diffraction measurements
were carried out on the High Intensity Powder Diffractometer at the
Manuel Lujan Neutron Scattering Center (MLNSC) at Los Alamos National
Laboratory and on the Glasses, Liquids and Amorphous Diffractometer at
the Intense Pulsed Neutron Source (IPNS) at Argonne National
Laboratory.  The samples were sealed in vanadium tubes with He
exchange gas. Data were collected as a function of temperature from
room temperature down to 10~K using a closed cycle He
refrigerator. Standard corrections~\cite{billi;prb93} were made to the
raw data, to account for experimental effects such as sample
absorption and multiple scattering, using the
program PDFgetN~\cite{peter;jac00}, to obtain the total scattering
structure function, $S(Q)$.  This contains both Bragg and diffuse
scattering and therefore information about atomic correlations on all
length scales.  The PDF, $G(r)$, is obtained by a Fourier
transformation according to $G(r) = {2\over\pi}\int_0^{\infty}
Q[S(Q)-1]\sin Qr\>dQ$, where Q is the magnitude of the scattering
vector. The PDF gives the probability of finding an atom at a distance
$r$ away from another atom.  The PDF from
La$_{1.875}$Sr$_{0.125}$CuO$_4$ at 300~K is shown in
Fig.~\ref{fig;data} (b)%
with the diffraction data in the form of $Q[S(Q)-1]$ in
Fig.~\ref{fig;data} (a). Superimposed on the PDF is a fit to the data
of the average structure model using the profile fitting
least-squares regression program, PDFFIT~\cite{proff;jac99}.  The
$S(Q)$ data were terminated at $Q_{max}=23$~\AA\sp{-1}.  This is a
conservative value for $Q_{max}$ in typical PDF measurements.  The
data from high-$Q$ has a poorer signal-to-noise ratio because of the
effect of the Debye-Waller factor.  By eliminating it from the Fourier
transform we improve the signal-to-noise ratio of our data and the
temperature to temperature reproducibility of the PDFs.  This reduces
the possibility that observed effects are noise artifacts.  We can
therefore have confidence that any effects that survive this
conservative approach to Fourier transforming the data are real.

\begin{figure}[tb]
\begin{center}$\,$
\epsfxsize=2.8in
\epsfbox{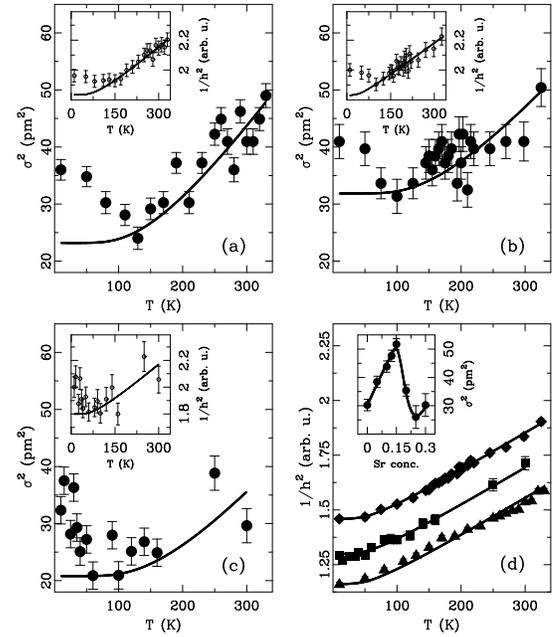}
\end{center}
\caption{Mean-square width, 
$\sigma^2$, of the in-plane Cu-O PDF peak at 1.9~\AA\ as a function of
temperature for (a) La$_{1.875}$Sr$_{0.125}$CuO$_4$, (b)
La$_{1.85}$Ba$_{0.15}$CuO$_4$, and (c) La$_{1.85}$Sr$_{0.15}$CuO$_4$. The
solid line gives the T-dependence predicted by the Einstein
model~\protect\cite{sevil;prb79}.  The insets show the inverse squared
peak height for the same peak with an Einstein curve superimposed.
(d) Temperature dependence of $1/h^2$ for the PDF peak at 7.2~\AA\
for La$_{1.875}$Sr$_{0.125}$CuO$_4$ (triangles),
La$_{1.85}$Ba$_{0.15}$CuO$_4$ (diamonds), and 
La$_{1.85}$Sr$_{0.15}$CuO$_4$ (squares) Inset shows the
strontium-doping dependence of $\sigma^2(x)$ for
La$_{2-x}$Sr$_x$CuO$_4$ at 10~K from Ref.~\protect\cite{bozin;prl00}.
}
\protect\label{fig;u2vt}
\end{figure}

In La$_{2-x}$(Sr,Ba)$_x$CuO$_4$, the first peak in the PDF at
$r=1.9$~\AA\ originates from the in-plane Cu-O bond. The width of this
peak comes from the relative motion of nearest neighbor in-plane Cu-O
pairs, plus any static or quasistatic bond-length distribution,
averaged over the whole sample. We have studied the mean square width,
$\sigma^2\propto\langle u^2\rangle$, of this peak as a function of
temperature for a series of underdoped La$_{2-x}$(Sr,Ba)$_x$CuO$_4$
compounds.  The results are reproduced in Fig.~\ref{fig;u2vt}.%
Peak profiles in the PDF are well modelled using a Gaussian function
convoluted with a Sinc function, $\sin Q_{max}
r/Q_{max}r$~\cite{billi;b;lsfd98}. The Sinc function results from
Fourier transforming the finite-range data. Since $Q_{max}$ is a known
experimental parameter it is possible to extract intrinsic peak widths
for the underlying Gaussian distributions.  The results of this
convoluted fitting process are shown in
Fig.~\ref{fig;u2vt}(a)-(c). It is clear from the Figure that the peak
width decreases with decreasing temperature as expected.  However,
below a certain temperature the Cu-O bond length distribution {\it
broadens} on further decrease of temperature.  This effect cannot be
explained by normal thermal behavior as indicated by the solid lines
which have the expected Einstein form~\cite{sevil;prb79}.  There also
are no structural phase transitions occuring at these temperatures.

The same qualitative result was obtained from the data directly
without carrying out a convoluted fit.  First, we simply plot the
inverse-squared PDF peak height, $1/h^2$, obtained directly from the
data.  This is a model independent measure of
$\sigma^2$ since the integrated area under the PDF peaks is
conserved~\cite{billi;b;lsfd98}.   The inverse-squared peak heights
 are shown in the insets to
Fig.~\ref{fig;u2vt}(a)-(c).  We also fitted the 1.9~\AA\ PDF peak
with pure Gaussian functions that were not convoluted with Sinc
functions (not shown).  Both these approaches reproduced the
qualitative result shown in Fig.~\ref{fig;u2vt}(a)-(c) giving us
confidence that it has a real origin and is not an artifact of the
convoluted fitting procedure.
All of these measures of the PDF peak width confirm the observation in
the convoluted peak fits that the underlying in-plane Cu-O pair
distribution is getting broader with decreasing temperature below some
temperature, $T_{si}$.

Peaks not involving in-plane Cu-O pairs, at higher values of $r$, in the
PDF behave canonically.  This is shown in Fig.~\ref{fig;u2vt}(d) where
$1/h^2(T)$ of the peak at $r=7.2$~\AA\ (indicated with an arrow in
Fig.~\ref{fig;data}) from each of the samples is plotted with an 
Einstein curve superimposed.  As expected, no upturn is observed at low
temperature.

The broadening of the $r=1.9$~\AA\ PDF peak at low temperature can be
explained if charge inhomogeneities, such as charge stripes, are
manifesting themselves in the structure at low temperature.  This will
occur both if the electronic correlations are appearing at low
temperature or if preexisting correlations are slowing down and
beginning to interact with the lattice.  It was shown in an earlier
PDF study~\cite{bozin;prl00} that a gradual broadening 
with increasing doping at 10~K of the
$r=1.9$~\AA\ PDF peak in
La$_{2-x}$Sr$_{x}$CuO$_4$ could be well explained
as a microscopic coexistence of heavily doped and undoped regions of
the copper-oxygen plane.  The $x$-dependence of this PDF peak width
measured at 10~K is reproduced in the inset to Fig.~\ref{fig;u2vt}(d).
This can be compared with the intrinsic peak widths at low temperature
from this study.

The original $x$-dependent data were interpreted as follows.  The
relatively sharp peaks in the $x=0,\ 0.25,$ and 0.30 data were assumed
to have a single valued bond length broadened by thermal and zero point motion
resulting in a mean-square width of $\sim 30$~pm\sp{2}.  The
relatively broader peaks observed in the underdoped compounds
($x=0.05$, 0.10, 0.125, 0.15) could be explained as a superposition of
sharp peaks that are shifted in position originating, respectively,
from less doped and more heavily doped regions of the copper oxide
plane~\cite{bozin;prl00}.  This very simple model independent analysis
is likely to be an oversimplification of the real situation where
local strains may lead to broader distributions of the PDF peaks;
however, it establishes unequivocably that the observed effects in the
PDF are consistent with structural distortions originating from charge
inhomogeneities.  Despite the current measurements being made on
different materials at different times using different diffractometers
it is clear that both the low-temperature thermal width of
25-31~pm\sp{2} extrapolated from the Einstein model, as well as the
excess peak height of $\sim 10-15$~pm\sp{2}, are in rather good
agreement with our earlier observation of the $x$-dependence of
La$_{2-x}$Sr$_{x}$CuO$_4$~\cite{bozin;prl00}.  This indicates that the
underlying origin of the peak broadening is the same.

The in-plane Cu-O pair correlation has been studied in a number of
XAFS measurements~\cite{haske;prb00,lanza;prl98,niemo;pc98}.  The data
of Lanzara {\it et al.}~\cite{lanza;prl98} qualitatively suggest an
upturn in the width of the distribution at low temperature.  However,
this result may not be significant since later work suggests that
uncertainties in unpolarized XAFS measurements are larger than the
observed effects ~\cite{niemo;pc98} and that polarized XAFS
measurements are necessary to obtain higher
precision~\cite{haske;prb00}.  In particular, this latter study puts
an upper limit of 0.017~\AA\ on possible non-thermal disorder amplitude
present in the in-plane Cu-O bond distribution of
La$_{1.875}$Ba$_{0.125}$CuO$_4$.  This is not far from our suggestion
of a $\sim 0.02$~\AA\ splitting observed in
La$_{2-x}$Sr$_{x}$CuO$_4$~\cite{bozin;prl00} and in the current work.
Our data will be compared with the result of Haskel {\it et al.} in
more detail elsewere~\cite{billi;ijmpb01;unpub}.  

\begin{figure}[tb]
\begin{center}$\,$
\epsfxsize=2.8in
\epsfbox{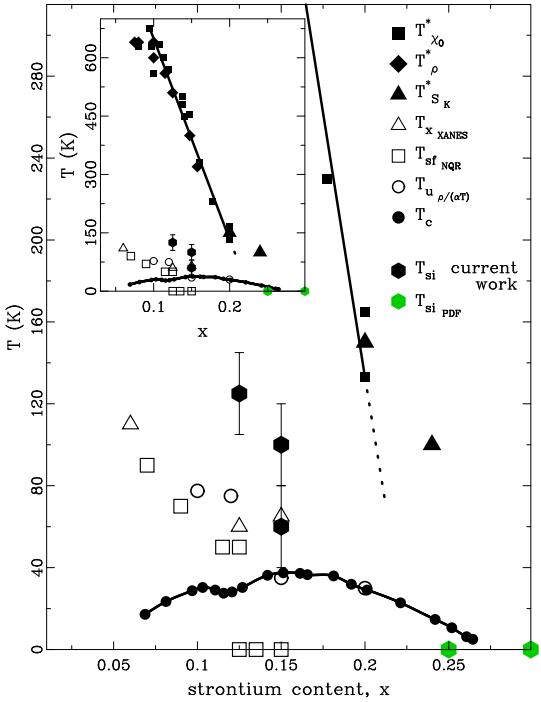}
\end{center}
\caption{Phase diagram of La$_{2-x}$Sr$_x$CuO$_4$ showing the 
temperatures of pseudogap opening , $T^*$~\protect\cite{timus;rpp99}, XANES
anomaly, $T_x$~\protect\cite{lanza;jpcm99}, NQR spin
freezing, $T_{sf}$~\protect\cite{hunt;prl99} transport
upturn, $T_u$~\protect\cite{ichik;cm99} and the $T_{si}$'s determined
from the present measurements.  $T_{si}$ is known to be below 10~K for
La$_{2-x}$Sr$_x$CuO$_4$ with $x>0.2$~\protect\cite{bozin;prl00} as
indicated.  $T_c$, is shown as solid circles joined by a line.  The
inset is the same phase diagram on an expanded temperature scale.}
\protect\label{fig;phased}
\end{figure}

We have extracted a temperature, $T_{si}$, where the structural
inhomogeneities set in by taking the difference, $\Delta\sigma^2$,
between the observed width and the Einstein curves plotted in
Fig.~\ref{fig;u2vt}.  The resulting values for $T_{si}$ are 125~K
for La$_{1.875}$Sr$_{0.125}$CuO$_4$ and 60~K and 100~K for
La$_{1.85}$Sr$_{0.15}$CuO$_4$ and La$_{1.85}$Ba$_{0.15}$CuO$_4$,
respectively. These are shown in Fig.~\ref{fig;phased} as solid
hexagons.  The estimated error bars are rather large since the exact
value of $T_{si}$ depends on parameters used in the Einstein
fits; also our data-sets are somewhat sparse.  However, they give a
temperature scale where the effects of charge inhomogeneities first
appear in the local structure.

In Figure~\ref{fig;phased} we show a phase diagram for
La$_{2-x}$Sr$_x$CuO$_4$ with T$_{si}$ plotted along with T$_c$
and T$^*$ obtained from the literature~\cite{timus;rpp99}.
 Superimposed on this diagram are 
T$_{sf}$, the onset temperature for NQR 
signal ``wipe-out''~\cite{hunt;prl99}, T$_u$, the temperature where the 
deviation of the normalized resistivity, $\rho/\alpha$T, 
reaches a critical value~\cite{ichik;cm99}, and T$_x$, the temperature 
where an anomaly is seen in XANES data~\cite{lanza;jpcm99}. 
All of these characteristic temperatures have been 
associated with charge or spin freezing. It is clear that the T$_{si}$'s 
obtained from the present data correlate quite well with the other measures 
of spin and charge freezing.

Our results clearly show that the charge inhomogeneities are strongly
coupled to the lattice in La$_{2-x}$(Sr,Ba)$_x$CuO$_4$ compounds and
become pinned by the lattice at low temperature. In the absence of Nd
the pinning is not complete and the charges do not order over long
range, even at x=0.125 in the Sr doped compound.  Nonetheless, their
strong coupling to the lattice will make them quite immobile.  Our
measurements yield the instantaneous structure and we cannot
distinguish whether the inhomogeneities are static or fluctuating on
phonon time scales or slower.  It will be interesting to see whether
similar effects are seen in the PDF of HgBa$_2$CuO$_{4+\delta}$ which
is a single layer cuprate superconductor like the
La$_{2-x}$(Sr,Ba)$_x$CuO$_4$ compounds but has a much higher T$_c$. It
is possible that electronically driven stripes are important for
superconductivity but a strong coupling to the lattice is destructive.
However, phonon anomalies have been associated with charge stripe
formation in YBa$_2$Cu$_3$O$_{6+\delta}$~\cite{mook;n99,petro;cm0300}
and theories exist in which the charge stripes are stabilized by the
lattice.  Resolving this issue will be a key component in gaining a
complete understanding of high temperature superconductivity.

We gratefully acknowledge D. G. Hinks, H. Takagi and G. H. Kwei for 
significant contributions as well as J. Johnson for her help with data
collection on GLAD. This work was supported by NSF through Grant No. 
DMR-0075149. MLNSC and IPNS are supported by the U.S. DOE under contracts 
No. W-7405-ENG-36 and No. W-31-109-ENG-38 respectively.

\bibliographystyle{/u24/billinge/bib/aip_simon}

\begin{thebibliography}{10}

\bibitem{timus;rpp99}
T.~Timusk and B.~Statt,
\newblock Rep. Prog. Phys. {\bf 62}, 61 (1999).

\bibitem{loese;s96}
A.~G. Loeser et al., 
\newblock Science {\bf 273}, 325 (1996).

\bibitem{homes;prl93}
C.~C. Homes, T.~Timusk, R.~Liang, D.~A. Bonn, and W.~N. Hardy,
\newblock Phys. Rev. Lett. {\bf 71}, 1645 (1993).

\bibitem{renne;prl98}
{Ch. Renner}, B.~Revaz, {J.-Y. Genoud}, K.~Kadowaki, and O.~Fischer,
\newblock Phys. Rev. Lett. {\bf 80}, 149 (1998).

\bibitem{batlo;pc94}
B.~Batlogg et al., 
\newblock Physica C {\bf 235-240}, 130 (1994).

\bibitem{warre;prl89}
J.~W.~W.~Warren et~al.,
\newblock Phys. Rev. Lett. {\bf 62}, 1193 (1989).

\bibitem{dai;prl96}
P.~Dai, M.~Yethiraj, H.~A. Mook, T.~B. Lindemer, and {F. Do\v gan},
\newblock Phys. Rev. Lett. {\bf 77}, 5425 (1996).

\bibitem{tranq;n95}
J.~M. Tranquada, B.~J. Sternlieb, J.~D. Axe, Y.~Nakamura, and S.~Uchida,
\newblock Nature {\bf 375}, 561 (1995).

\bibitem{yamad;prb98}
K.~Yamada et al., 
\newblock Phys. Rev. B {\bf 57}, 6165 (1998).

\bibitem{mook;n99}
H.~A. Mook and {F. Do\v gan},
\newblock Nature {\bf 401}, 145 (1999).

\bibitem{schul;prl89}
H.~J. Schulz,
\newblock Phys. Rev. Lett. {\bf 64}, 1445 (1989).

\bibitem{emery;pnas99}
V.~J. Emery, S.~A. Kivelson, and J.~M. Tranquada,
\newblock Proc. Natl. Acad. Sci. USA {\bf 96}, 8814 (1999).

\bibitem{caste;zpb97}
C.~Castellani, {C. DiCastro}, and M.~Grilli,
\newblock Z. Phys. B {\bf 103}, 137 (1997).

\bibitem{bianc;ssc97}
A.~Bianconi, A.~Valletta, A.~Perali, and N.~L. Siani,
\newblock Solid State Commun. {\bf 102}, 369 (1997).

\bibitem{white;prl98}
S.~R. White and D.~J. Scalapino,
\newblock Phys. Rev. Lett. {\bf 80}, 1272 (1998).

\bibitem{marti;cm0300}
I.~Martin, G.~Ortiz, A.~V. Balatsky, and A.~R. Bishop,
\newblock cond-mat/0003316.

\bibitem{phill;pmb99}
J.~C. Phillips,
\newblock Philos. Mag. B {\bf 79}, 527 (1999).

\bibitem{marki;jpcs97}
R.~S. Markiewicz,
\newblock J. Phys. Chem. solids {\bf 58}, 1179 (1997).

\bibitem{ichik;cm99}
N.~Ichikawa et al., 
\newblock cond-mat/9910037.

\bibitem{mook;n00}
H.~A. Mook, P.~Dai, F.~Do\v{g}an, and R.~D. Hunt,
\newblock Nature {\bf 404}, 729 (2000).

\bibitem{hunt;prl99}
A.~W. Hunt, P.~M. Singer, K.~R. Thurber, and T.~Imai,
\newblock Phys. Rev. Lett. {\bf 82}, 4300 (1999).

\bibitem{lanza;jpcm99}
A.~Lanzara et al., 
\newblock J. Phys: Condens. Matter {\bf 11}, L541 (1999).

\bibitem{petro;cm0300}
Y.~Petrov et al., 
\newblock cond-mat/0003414.

\bibitem{egami;b;lsfd98}
T.~Egami,
\newblock in {\em Local Structure from Diffraction}, edited by S.~J.~L.
  Billinge and M.~F. Thorpe, page~1, New York, 1998, Plenum.

\bibitem{bozin;prl00}
E.~S. Bo{\v z}in, S.~J.~L. Billinge, H.~Takagi, and G.~H. Kwei,
\newblock Phys. Rev. Lett. {\bf 84}, 5856 (2000).

\bibitem{radae;prb94i}
P.~G. Radaelli et al., 
\newblock Phys. Rev. B {\bf 49}, 4163 (1994).

\bibitem{goode;f92}
J.~B. Goodenough,
\newblock Ferroelectrics {\bf 130}, 77 (1992).

\bibitem{petko;prl99}
V.~Petkov et al., 
\newblock Phys. Rev. Lett. {\bf 83}, 4089 (1999).

\bibitem{billi;prl93}
S.~J.~L. Billinge, G.~H. Kwei, A.~C. Lawson, J.~D. Thompson, and H.~Takagi,
\newblock Phys. Rev. Lett. {\bf 71}, 1903 (1993).

\bibitem{breue;pc93}
M.~Breuer et al., 
\newblock Physica C {\bf 208}, 217 (1993).

\bibitem{billi;prb93}
S.~J.~L. Billinge and T.~Egami,
\newblock Phys. Rev. B {\bf 47}, 14386 (1993).

\bibitem{peter;jac00}
P.~F. Peterson, M.~Gutmann, {Th.~Proffen}, and S.~J.~L. Billinge,
\newblock J. Appl. Crystallogr. {\bf 33}, 1192 (2000),
\newblock (copyright International Union of Crystallography).

\bibitem{proff;jac99}
{Th. Proffen} and S.~J.~L. Billinge,
\newblock J. Appl. Crystallogr. {\bf 32}, 572 (1999).

\bibitem{sevil;prb79}
E.~Sevillano, H.~Meuth, and J.~J. Rehr,
\newblock Phys. Rev. B {\bf 20}, 4908 (1979).

\bibitem{billi;b;lsfd98}
S.~J.~L. Billinge,
\newblock in {\em Local Structure from Diffraction}, edited by S.~J.~L.
  Billinge and M.~F. Thorpe, page 137, New York, 1998, Plenum.

\bibitem{haske;prb00}
D.~Haskel, E.~A. Stern, F.~Dogan, and A.~R. Moodenbaugh,
\newblock Phys. Rev. B {\bf 61}, 7055 (2000).

\bibitem{lanza;prl98}
A.~Lanzara et al., 
\newblock Phys. Rev. Lett. {\bf 81}, 878 (1998).

\bibitem{niemo;pc98}
T.~Niem{\" o}ller et al., 
\newblock Physica C {\bf 299}, 191 (1998).

\bibitem{billi;ijmpb01;unpub}
S.~J.~L. Billinge, M.~Gutmann, and {E. S. Bo\v zin},
\newblock (2000),
\newblock unpublished.

\end{thebibliography}

\end{document}